# CHARACTERIZATION OF THE BEAM FROM THE RFQ OF THE PIP-II INJECTOR TEST*

A. Shemyakin[†], J.-P. Carneiro, B. Hanna, L. Prost, A. Saini, V. Scarpine, V.L.S. Sista[1], J. Steimel,
Fermilab, Batavia, IL 60510, USA
[1]also at Bhabha Atomic Research Centre (BARC), Mumbai, India

## Abstract

A 2.1 MeV, 10 mA CW RFQ has been installed and commissioned at the Fermilab's test accelerator known as PIP-II Injector Test. This report describes the measurements of the beam properties after acceleration in the RFQ, including the energy and emittance.

## INTRODUCTION

The PIP-II Injector Test (PIP2IT) [1] is a test accelerator being assembled at Fermilab to address critical issues associated with the low-energy part of a future CW-compatible $H^-$ superconducting linear accelerator PIP-II [2]. Its warm front end includes a 10 mA DC, 30 keV $H^-$ ion source, a Low Energy Beam Transport (LEBT) with a chopper, a 2.1 MeV CW RFQ, followed by a Medium Energy Beam Transport (MEBT). The ion source, LEBT, RFQ, and short MEBT with beam diagnostics were commissioned [3]. In subsequent sections we describe the results from measuring the RFQ beam transmission, output energy and beam emittance.

## PIP2IT RFQ

The PIP2IT RFQ is a 4-vane, 4-section CW RFQ designed and constructed at LBNL [4]. Some of the RFQ specifications are listed in Table 1.

Table 1: RFQ Specifications

| Parameter | Value |
|---|---|
| Input Energy | 30 keV |
| Output Energy | 2.1 MeV +/- 1% |
| Frequency | 162.5 MHz |
| Beam Current, nominal( max) | 5 (10) mA |
| Nominal Vane Voltage | 60 kV |
| RF Power | $\leq$ 130 kW |
| Transmission | $\geq$ 95% |
| Transverse Emittance | $\leq$ 0.2 μm |
| Bunch Longitudinal Emittance (at 5 mA) | $\leq$ 0.28 μm (0.88 eV-μs) |

Prior to measurements with beam, the RFQ was conditioned to full power with both pulsed and CW RF [5].

## MEASUREMENT SETUP

The beam is injected into the RFQ in pulses with length of 10 μs to 8 ms at a repetition frequency ranging from single bunches to 60 Hz, formed by the LEBT chopper.

The beam from the ion source can be either DC or pulsed at the same frequency as the chopper. In the latter case, the chopper passes the beam into the RFQ 1-2 ms after the beginning of the ion source pulse to allow beam neutralization near the ion source to reach a steady state. To transport the beam through the MEBT diagnostics, the beam is focused transversely by two quadrupole doublets (manufactured by BARC, India) and longitudinally by a 162.5 MHz bunching cavity. Each doublet is followed by a two-plane dipole corrector. Instruments in this focusing section include a Pearson 7655 AC current transformer (a.k.a. toroid); Beam Position Monitors (BPM) [6], a capacitive pickup attached to the upstream quadrupole of each doublet; and a set of 4 scrapers mounted in one vacuum enclosure. Each scraper is an electrically isolated, radiation-cooled, independently movable plate made of the molybdenum alloy TZM, and capable of intercepting up to 75 W of average power [7].

Various instruments were installed immediately downstream of the quadrupole doublets in several different configurations of the MEBT for RFQ beam characterization. One of the configurations, shown in Fig. 1, includes one more set of scrapers and a toroid, both identical to the corresponding components upstream, and an emittance scanner used for recording the beam phase portrait in the horizontal plane (shown in vertical orientation in Fig.1 for presentation purpose).

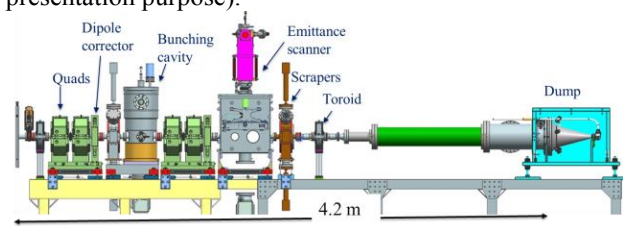

Figure 1: One of the beam line configurations for characterization of the RFQ beam.

The Allison-type emittance scanner is similar to the one installed in the LEBT [9] and shares electronics and software with it. The phase portraits along a pulse can be recorded in time bins of length down to 1 μs. The beam passed through the diagnostics is delivered into a beam dump supplied by Oak Ridge National Lab (ORNL) [8].

Two other devices installed in alternate configurations, a Time-of-Flight (ToF) monitor and a Fast Faraday Cup (FFC), are described in the next section.

## BEAM MEASUREMENTS

Most of the RFQ beam specifications were successfully verified.

Transmission through the RFQ is measured as a difference between currents read by three identical toroids on both sides of the RFQ. The toroids are periodically calibrated with the same source, and the drifts are found to be less than 2%. For the LEBT best tune and nominal RFQ voltage of 60 kV, the measured transmission is 98%. Simulations of the RFQ initialized with the distribution measured at the end of the LEBT predict that all but ~1% of the particles transmitted through the RFQ are accelerated. The rest exits the RFQ as DC beam with the energy of 30-40 keV. Such beam is nearly eliminated in the field of the MEBT magnets tuned for transporting the 2.1 MeV beam. Therefore, the amount of the DC beam can be estimated by comparing the signals from the toroids located downstream of the RFQ and downstream of the second doublet. The signals are found identical implying that the DC beam current is below the measurement accuracy of 2%.

The energy of the ions coming out of the RFQ is measured with the ToF monitor, which idea originated from SNS [10]. It is a capacitive pickup mounted on a platform that can be moved precisely by 25 mm along the beam line axis (Fig. 2a).

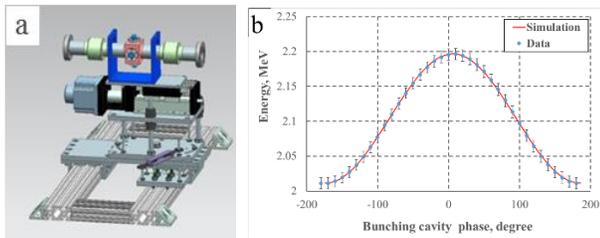

Figure 2: ToF monitor model (**a**) and measurement of the energy as a function of the bunching cavity phase (**b**).

The phase of the ToF pickup signal is recorded as a function of the pickup longitudinal position, and the slope of the line is inversely proportional to the beam velocity. Also, the ToF monitor is used for calibration of the voltage and phase of the bunching cavity (Fig. 2b). Each point represents fitting of 720 phase/position measurements with statistical error of 0.2 - 0.3%. The systematic error, determined mainly by the accuracy of the mechanical motion, is significantly lower, ~0.1%.

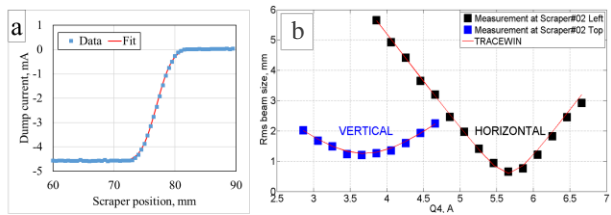

Figure 3. Beam size measurements with scrapers. **a**- currents of the dump as a function of the horizontal scraper position. The red line is a fit assuming the Gaussian beam distribution with rms width of 1.97 mm. Each data point is the average of ten 10-μs pulses. **b**- example of a quadrupole scan. Fitted emittances are 0.17 μm horizontal and 0.16 μm vertical (rms n). The data point at the quadrupole current of 5.06 A is deduced from the plot in (a).

The beam size is measured by moving the scrapers, one at a time, through the beam and recording the current downstream with the toroid or beam dump (Fig. 3a). The data are fitted to an integral function that assumes that the beam has a 2D Gaussian distribution. The resulting rms sizes are reproducible to < 0.2 mm (with scatter of 0.02 mm rms) when the measurements are repeated for several times or made with opposing scrapers (e.g. by the left and right scrapers in the same set). The scraper motion was preliminary calibrated to 1% accuracy.

To calculate the transverse emittance, the beam sizes are measured at the second scraper set while changing the current of the last quadrupole (Fig. 3b). Then the Twiss parameters at the quadrupole and emittance are calculated in zero space charge approximation, backpropagated to the RFQ exit, and used as the initial guess for fitting with TraceWin simulations [11] that include space charge. With optimal tuning and the nominal beam current of 5 mA, the emittance is measured below the specification of 0.2 μm (rms n) in both planes.

In addition to quadrupole scans, the emittance in the horizontal plane was measured with the Allison scanner. An example of a measured distribution is shown in Fig. 4a. The values of the rms emittance measured with the scanner and with the quadrupole scans are within estimated measurement errors (< 10 %) (Fig. 4b).

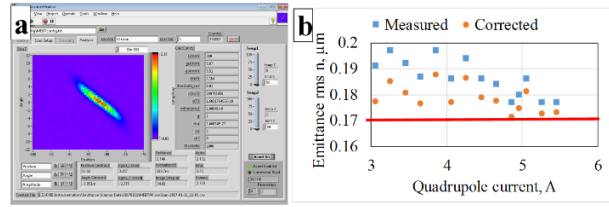

Figure 4: Allison scanner measurements. **a**- screenshot of a phase portrait recorded with a 10 μs, 5 mA pulse. Time bin is 5 μs. For the calculations, the background cut is 1% of the maximum intensity. **b**- comparison of the rms horizontal emittance measured with the quadrupole scan (red line) and with the Allison scanner for different currents of the most downstream quadrupole. The blue rectangles represent measured data, and the orange circles are the same data corrected for the effect of the finite slits size (0.2 mm) with formulae from [12]. The portrait in (a) was recorded at the quadrupole current of 4.66 A.

While most of the Allison scanner measurements were done with 10-20 μs pulses to avoid damaging its front slit, the scanner is capable of handling up to 0.5 ms beam pulses with a large beam size. Initially the long pulse beam parameters were found to be varying at the 10-15% level. It was traced to peculiarities of the neutralization process in the LEBT. After optimizing the potentials of the LEBT electrodes controlling neutralization and running DC beam out of the ion source, the Twiss parameters and emittance became flat within the measurement scatter, ≤ 5% (Fig. 5). In this measurement, the LEBT focusing was not fully optimized so that the emittance is slightly higher than in Fig. 4.

The beam current as measured by the image integral of the phase space portrait drops monotonically by 4% throughout the pulse. The reason for this drop is not clear.

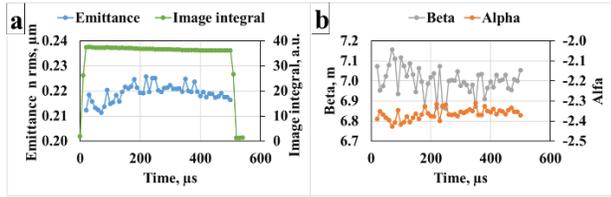

Figure 5: Parameters variations throughout a 0.5 ms pulse (10 µs time bins). Except for the image integral, other data are shown with suppressed zero and only for points with full image integral. **a-** emittance and image integral, **b-** Twiss functions in horizontal plane.

Another effect not properly understood is the motion of the beam center during the pulse. A ~1 mm displacement is observed by the BPMs and confirmed in the scanner measurements. One of the explanations considered is related to the beam trajectory variation with changes of the RFQ voltage reported in Ref. [13]. However, the measured flatness of the RFQ RF pulse makes a noticeable contribution of this effect unlikely. Another possibility may be the deflection of the beam in the LEBT caused by the chopper transition process.

The bunch length is measured with a Fast Faraday Cup (FFC). The 0.8 mm hole in the insertable head of the FFC cuts out a beamlet that is absorbed by a collector separated from the grounded surface by 1.7 mm. The collector signal, digitized by a 4-GHz oscilloscope (Fig. 6a), should represent the bunch shape accurately (<10%) down to a bunch rms width of 0.1 ns. While the measured bunch length behaves similarly to the simulations, as shown in Fig. 6b, the amplitude of the recorded FFC current is several times lower than expected, and we are not yet ready to claim that the longitudinal emittance has been accurately measured.

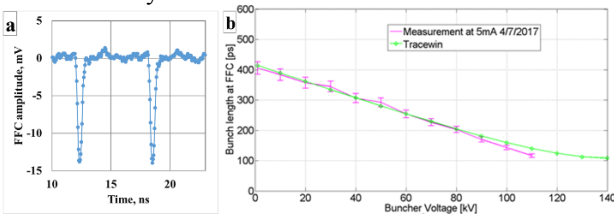

Figure 6: FFC measurements. **a**- scope trace for two bunches in the middle of a 10-µs pulse. **b**- bunch length as a function of the bunching cavity voltage. The cavity phase is at -90°. Pink points are data, and the green curve is the result of simulations with TraceWin for the rms emittance of 0.22 µm. The trace in (**a**) is recorded at 90 kV.

The measurements discussed above were performed at the nominal RFQ inter-vane voltage of 60 kV. Decrease of the voltage below ~58 kV results in a significant increase of the transverse emittance, a measurable drop of the ion energy and RFQ transmission, and deviation of Twiss functions (Fig. 7).

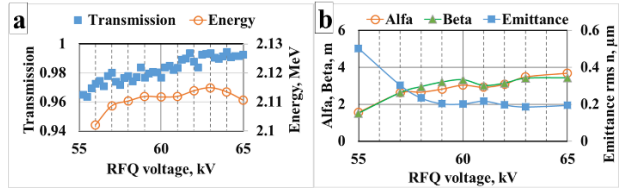

Figure 7. Beam parameters as functions of the RFQ voltage.

## SUMMARY

The beam accelerated in the PIP2IT RFQ is fully characterized in the transverse planes and satisfies the PIP-II requirements. Longitudinal measurements have been performed but not at a satisfactory level yet. When the neutralization process is properly controlled in the LEBT, the beam parameters are constant in the MEBT throughout a 0.5 ms pulse. Beam position variations during long pulses need to be addressed.

## ACKNOWLEDGMENT

Authors are thankful to the entire PIP2IT team who made the measurements possible. Particularly, we would like to acknowledge the work of M. Alvarez, R. Andrews, C. Baffes, A. Chen, P. Derwent, J. Edelen, B. Fellenz, B. Hartsell, K. Kendziora, M. Kucera, D. Lambert, D. Petersen, A. Saewert, G. Saewert, D. Sun, and A. Warner.